\documentclass[onecolumn]{emulateapj}

\submitted{ApJ, in press}

\shorttitle{Massive disks in triaxial halos}
\shortauthors{Bailin et~al.}

\defcitealias{jog00}{J2K}
\defcitealias{hn06}{HN06}
\defcitealias{hns07}{HNS}

\newcommand{\Msun}{{\ensuremath{\mathrm{M}_{\sun}}}}
\newcommand{\fpert}{\ensuremath{f_{\mathrm{pert}}}}
\newcommand{\epsdisk}{{\ensuremath{\epsilon_{\mathrm{disk}}}}}
\newcommand{\diffd}{\ensuremath{\mathrm{d}}}
\newcommand{\vesc}{\ensuremath{V_{\mathrm{esc}}}}

\begin{document}

\title{Self-consistent massive disks in
 triaxial dark matter halos}

\author{Jeremy Bailin\altaffilmark{1,2}, Joshua D. Simon\altaffilmark{3},
Alberto D. Bolatto\altaffilmark{4}, Brad K. Gibson\altaffilmark{5}
and Chris Power\altaffilmark{1}}
\email{jbailin@astro.swin.edu.au}
\altaffiltext{1}{Centre for Astrophysics and Supercomputing,
 Swinburne University of Technology, Mail H39, PO Box 218, Hawthorn, Victoria,
 3122, Australia}
\altaffiltext{2}{Current address: Department of Physics \&\ Astronomy, ABB-241,
 McMaster University, 1280 Main St. W, Hamilton, ON L8S 4M1, Canada}
\altaffiltext{3}{Department of Astronomy, California Institute of Technology,
 1200 East California Boulevard, MS 105-24, Pasadena, CA 91125, USA}
\altaffiltext{4}{Department of Astronomy and Radio Astronomy Laboratory,
 University of California at Berkeley, Berkeley, CA 94720, USA}
\altaffiltext{5}{Centre for Astrophysics, University of Central Lancashire,
 Preston, PR1 2HE, UK}

\begin{abstract}
Galactic disks in triaxial dark matter halos become deformed
by the elliptical potential in the plane of the disk in such a way as to
counteract the halo ellipticity.
We develop a technique to calculate the equilibrium configuration
of such a disk in the combined disk-halo potential, which is
based on the method of \citet{jog00} but accounts
for the radial variation in both the halo potential
and the disk ellipticity.
This crucial ingredient results in qualitatively
different behavior of the disk: the disk circularizes
the potential at small radii, even for a
reasonably low disk mass. This effect has important implications
for proposals to reconcile cuspy
halo density profiles with low surface brightness galaxy
rotation curves using halo triaxiality.
The disk ellipticities in our models are consistent with
observational estimates based on two-dimensional
velocity fields and isophotal axis ratios.
\end{abstract}


\section{Introduction}\label{intro section}
Galaxies are thought to be surrounded by large dark matter halos.
These halos are much more massive than the visible components
of galaxies, and dominate much of the dynamics.
Although dark matter halos are often assumed to be spherical
for simplicity,
the halos that form in
cosmological simulations
are quite flattened, with typical intermediate axis ratios
of $b/a \sim 0.8$ and minor axis ratios of $c/a \sim 0.6$,
with some systematic
variation depending on the mass of the halo and the
radius at which the shape is measured
\citep{warren-etal92,js02,bs05-alignment,allgood-etal06}.
This non-sphericity is a testable prediction of
cosmological models.

Simulations of disk galaxy formation within dark matter halos
find that the presence of the disk modifies the shape
of the halo, reducing the halo triaxiality 
\citep{dubinski94,kazantzidis-etal04,bailin-etal05-diskhalo,bs06}.
However,
as long as the final shape of the halo retains some ellipticity in the
plane of the disk, the dynamics
and shape of the disk will be affected by
the deviations from axisymmetry \citep[e.g.][]{gv86,sfdz97}.

Observations indicate that many disks
do indeed have small but non-zero
ellipticities. Evidence for elliptical disks
comes from harmonic decomposition of galaxy photometry
\citep{rz95}, harmonic decomposition
of two-dimensional velocity fields \citep{sfdz97,simon-etal05},
and statistical analysis of the distribution of projected
shapes \citep{ryden06}.
These results qualitatively confirm that galactic dark matter
halos are elliptical; precision measurements could provide
direct constraints on the shapes of the halos.

Recently, \citet[][hereafter HN06]{hn06} proposed that elliptical orbits within
the disk produced by the triaxiality of the halo could reconcile
cuspy density profiles that form in cosmological simulations
\citep[][hereafter NFW]{nfw96} with observed rotation curves
of low surface brightness (LSB) galaxies, which often
appear to require halos with constant density cores
\citep[e.g.,][]{deblok-etal01}.
This analysis did not take into account the self-gravity of the disk.  In
galaxies with massive disks, the gravity of the disk contributes to
the net potential and the dynamics of the disk are determined by a
combination of the halo and the disk itself. In order to draw
conclusions about the shape of the halo from the measured dynamics of
the disk, we must determine self-consistently both how the disk is
perturbed by the potential and how the perturbed disk contributes to
the potential.

An elegant method to carry out these calculations was proposed by \citet{jog00} (hereafter
\citetalias{jog00}; see also \citealt{jog97,jog99}). By assuming a
logarithmic halo potential with a small constant elliptical perturbation
and an exponential disk with a small constant elliptical response,
\citetalias{jog00} solved for the self-consistent response. She
demonstrated that the disk response dilutes the ellipticity
of the potential most strongly at $1.42$ disk scale lengths.

There are a number of simplifying assumptions in \citetalias{jog00} that
require examination. The most important
assumption is that both the halo perturbation and the disk response
are constant with radius.
In contrast, \citetalias{hn06} demonstrated
that a radially-varying perturbation is required to reconcile LSB long-slit
rotation curves with cuspy halo profiles.
Indeed, cosmological simulations predict a radially-varying
perturbation in the halo potential;
even if halos had isodensity surfaces of constant ellipticity,
the shape of the potential would vary with radius,
and \citet[][hereafter HNS]{hns07},
who directly measured the shapes of isopotential surfaces
of cosmological halos,
found even stronger variation with radius.
The response of the disk is also not expected to be uniform
in a radially-varying potential.

In this paper, we generalize the method of \citetalias{jog00} to the 
more realistic case of radially varying halo perturbations and radially
varying disk responses. In \S~\ref{method section} we detail the method
for determining the disk shape and dynamics. In \S~\ref{results section}
we use this method to determine the shapes and dynamics of disks in
sample triaxial halos and demonstrate how the results depend
on the properties of the disks and halos.
\S~\ref{discussion section} discusses our results in the
context of observations that directly probe disk ellipticity, and in
\S~\ref{conclusions section} we present our conclusions.

\section{Method}\label{method section}

\subsection{Outline}\label{outline method section}

Our method, which is based closely on \citetalias{jog00},
is as follows:
\begin{enumerate}
 \item Calculate the axisymmetric component of the potential and the
	elliptical perturbation in the potential induced by the triaxial halo
	(\S~\ref{axisymmetric potential section}).
 \item Calculate the closed orbits and corresponding disk ellipticity
	for a given net perturbation to the potential
	(\S\S~\ref{closed orbit section} and \ref{disk ellipticity section}).
 \item Calculate the elliptical perturbation in the potential induced by
	a given disk ellipticity (\S~\ref{disk potential section}).
 \item Solve for the form of the net potential perturbation that satisfies
	all of the above constraints (\S~\ref{self-con soln section}).
\end{enumerate}
Throughout this procedure, the halo is kept fixed, i.e. it does not
respond to the presence of the disk. Therefore, the potential that should be
used in the calculation is the real shape of the halo in which
the disk lies, which is less triaxial than the shape the halo would
have in the absence of baryonic processes
\citep{kazantzidis-etal04,bailin-etal05-diskhalo,bs06}.

The main differences between this work and \citetalias{jog00} are:
\begin{enumerate}
 \item Where \citetalias{jog00} assumes a logarithmic potential for the halo,
	we evaluate the radial form of the potential directly from
	a density distribution motivated by cosmological simulations.
 \item Where \citetalias{jog00} assumes a constant perturbation
	to the halo potential, we allow the perturbation to vary with
	radius and either evaluate it directly from a triaxial density
	distribution motivated by cosmological simulations or 
	use parametrizations developed from measurements of halos
	in cosmological simulations.
 \item Where \citetalias{jog00} assumes that the disk responds with a constant
	ellipticity, we allow the disk ellipticity to vary with radius.
\end{enumerate}
Whenever we carry out numerical calculations in this paper,
all radially-varying functions are tabulated on a radial grid
sampled at $100$ radii $R_i$ spaced logarithmically between $0.1$
and $100$~kpc in order to finely sample the inner region of the disk
where the quantities vary most rapidly; $50$ grid points lie at
$R<1$~kpc. The functions are linearly interpolated
between grid points when their values are required at arbitrary
radii.

\subsection{Axisymmetric potential and halo perturbation}%
\label{axisymmetric potential section}
For the perturbative approach, we assume that within the plane of the disk
(which we take to be $z=0$ for simplicity), the total potential can be written as
\begin{eqnarray}
  \Phi(R,\theta) &=& \Phi_0(R) + \Phi_{\mathrm{pert}}(R,\theta)\nonumber\\
   &=& \Phi_0(R) \left(1 + \fpert(R) \cos m\theta\right),
\label{eq fpert defn}
\end{eqnarray}
where $R$, $\theta$, and $z$ are the cylindrical coordinates.
For the purposes of this paper, we will assume an elliptical
perturbation (i.e., $m=2$) from now on
(see Appendix~\ref{second order terms section} for a discussion of
the $m=4$ mode).
We assume that $\fpert(R)$ is small and varies slowly with $R$.
Note that for $\fpert > 0$, isopotentials are elongated along the $x$ axis
and closed orbits in the disk are elongated along the $y$ axis.

Both the disk and halo contribute to both the axisymmetric and $m=2$ components
of the potential:
\begin{equation}
  \Phi_0(R) = \Phi_0^{\mathrm{halo}}(R) + \Phi_0^{\mathrm{disk}}(R),
\end{equation}
\begin{equation}
  \fpert(R) = \fpert^{\mathrm{halo}}(R) + \fpert^{\mathrm{disk}}(R).
\label{eq fpert breakdown}
\end{equation}

To first order,
the $m=2$ perturbation in the potential induces
an $m=2$ perturbation in the surface density distribution of the
otherwise exponential disk
(see Appendix~\ref{second order terms section} for a justification of
our decision to neglect the higher-order terms):
\begin{equation}
  \Sigma(R,\theta) = \Sigma_0 \exp \left[ - \frac{R}{R_d}
    \left( 1 - \frac{\epsdisk(R)}{2} \cos 2\theta \right) \right]
 \label{disk density distribution}
\end{equation}
We assume $\epsdisk(R)$, the ellipticity of the isodensity ellipse,
is small and varies slowly with $R$.
The axisymmetric component of the disk potential is given by
\begin{equation}
  \Phi_0^{\mathrm{disk}}(R) = - \pi G \Sigma_0 R \left[ I_0(y)
    K_1(y) - I_1(y) K_0(y) \right]
\end{equation}
($y=R/2 R_d$; see \citealt{freeman70}; \citealt{BT}, eq. 2-168), where
$I$ and $K$ are modified Bessel functions.

Given the density distribution of the dark matter halo,
both axisymmetric and $m=2$ components of the halo potential can be
evaluated.
If the isodensity surfaces of the halo are self-similar ellipsoids,
then the halo density can be written as:
\begin{equation}
  \rho(x,y,z) = \rho(s),
\end{equation}
where
\begin{equation}
  s^2 = \left(\frac{x}{a}\right)^2 + \left(\frac{y}{b}\right)^2 + 
    \left(\frac{z}{c}\right)^2.
\end{equation}
For example, we can use an NFW form for the density:
\begin{equation}
  \rho(s) = \frac{\rho_0}{\frac{s}{r_s}\left(1 + \frac{s}{r_s}\right)^2}
\label{NFW density eq}
\end{equation}
\citep{nfw96,js02}.
We make no assumptions about the relative magnitudes
of $a$, $b$, and $c$; they are simply the relative axis ratios along the $x$,
$y$, and $z$ axes respectively. Therefore, the disk,
which lies in the $xy$ plane, can be oriented in
any of the principal planes of the halo.

We calculate the halo potential along the $x$ axis,
$\Phi_x^{\mathrm{halo}}(R) \equiv \Phi^{\mathrm{halo}}(R,0,0)$,
and along the $y$ axis,
$\Phi_y^{\mathrm{halo}}(R) \equiv \Phi^{\mathrm{halo}}(0,R,0)$,
by numerically integrating
eq. (2-99) of \citet{BT} using $\rho(s)$ 
given in equation (\ref{NFW density eq}).
This allows us to calculate $\Phi_0^{\mathrm{halo}}$ and
$\fpert^{\mathrm{halo}}$ as 
\begin{equation}
  \Phi_0^{\mathrm{halo}}(R) = \frac{1}{2} \left[\Phi_x(R) + \Phi_y(R)\right],
\end{equation}
and
\begin{equation}
  \fpert^{\mathrm{halo}}(R) =
    \frac{1}{2 \Phi_0(R)} \left[\Phi_x(R) - \Phi_y(R)\right].
\label{eq fperthalo}
\end{equation}

\subsection{Closed orbits}\label{closed orbit section}
In a triaxial potential,
dissipative gas settles on stable closed loop orbits when
such orbits exist \citep{elzant-01}. This is the case throughout
the potential of a centrally-concentrated mass profile such
as the NFW profile. Therefore, the structure of a galactic
disk, which consists of gas clouds and stars formed within those gas clouds,
is determined by the form of the closed orbits.
These have been examined
in detail by \citet{sfdz97}, and simplified into a convenient
form by \citetalias{hn06}.
These previous derivations have assumed that the perturbation to the
potential is constant over the radial excursion of an orbit, which is
not the case for the radially-varying perturbations that we wish to
study. We have therefore rederived the equations for closed orbits
within a radially-varying perturbation from the equations of motion.
The orbits follow
\begin{equation}
 R = R_0 \left( 1 - \frac{\fpert\, a_{12}}{2}
   \cos 2 \theta_0 \right)
\label{closed orbit R}
\end{equation}
\begin{equation}
  \theta = \theta_0 + \frac{a_{12}+a_{32}}{2m}
    \fpert \sin 2 \theta_0,
\label{closed orbit theta}
\end{equation}
with velocities
\begin{equation}
  V_R = V_c\, \fpert\, a_{12} \sin 2 \theta_0 \label{vr defn}
\label{closed orbit VR}
\end{equation}
\begin{equation}
  V_{\theta} = V_c \left( 1 + \frac{\fpert\, a_{32}}{2} \cos 2 \theta_0
    \right), \label{vtheta defn}
\label{closed orbit Vtheta}
\end{equation}
where $R_0$ and $\theta_0$ define the guiding center of the orbit,
$\theta_0=\Omega_0\, t$, and the following
functions of $\Phi_0(R)$ are evaluated at $R_0$:
\begin{equation}
 \Omega_0(R) = \sqrt{\frac{1}{R} \frac{\diffd \Phi_0}{\diffd R}}
\end{equation}
\begin{equation}
 V_c(R) = R\, \Omega_0(R)
\end{equation}
\begin{equation}
 \vesc(R) = \sqrt{2 |\Phi_0|}
\end{equation}
\begin{equation}
 g_m(R) = \frac{1}{\Omega_0^2} \frac{\diffd^2 \Phi_0}{\diffd R^2} - (m^2 - 3)
\end{equation}
\begin{equation}
 a_{1m}(R) = \frac{2}{g_m(R)}\left[ 1 - \frac{\vesc^2}{V_c^2}
  \left(1 + \frac{1}{2} \frac{R}{\fpert} \frac{\diffd \fpert}{\diffd R}\right)
  \right]
\label{a1m definition}
\end{equation}
\begin{equation}
 a_{3m}(R) = a_{1m}(R) + \frac{\vesc^2}{V_c^2}.
\end{equation}
The coefficients $a_{1m}$ and $a_{3m}$ quantify the degree
to which the radius $R$ and angular velocity $V_{\theta}$ respectively,
which are constant for a circular orbit, vary for a unit
perturbation to the potential.
Our expressions differ from those in \citetalias{hn06} for the following
reasons:
(1) we have taken the radial variation of the perturbation into account
in our derivation of equation (\ref{a1m definition});
(2) we have generalized the expression for $g_m(R)$ to be valid for
all $m$, while the expression in \citetalias{hn06} is specific to $m=2$;
(3) we have removed the factor of \fpert\ from the definitions of $a_{1m}$
and $a_{3m}$ for convenience later (note that these quantities still
depend implicitly on \fpert\ through its derivative);
and (4) we have generalized the expression for $a_{3m}$ to be valid
for all potential profiles, while the expression in \citetalias{hn06} is
only valid when $g_2(R)=-3$, which is not the case in the inner
regions of an NFW potential.

Given the tabulated values of $\Phi_0$, these functions can be evaluated
at the grid points $R_i$. Because $\Phi_0$ has been calculated from
analytic functions, the tabulated values are relatively free of
noise and even the numerical second derivative does not contain large
fluctuations.
Note that the closed orbits are elongated \emph{perpendicular} to the
isopotential contours.

\subsection{Disk ellipticity}\label{disk ellipticity section}
The disk must satisfy the continuity equation. In cylindrical
coordinates:
\begin{equation}
\frac{\partial}{\partial R} \left[ R\, \Sigma(R,\theta)\,
  V_R(R,\theta) \right] + 
\frac{\partial}{\partial\theta} \left[ \Sigma(R,\theta)\,
  V_{\theta}(R,\theta) \right] = 0.
\end{equation}
We substitute $\Sigma(R,\theta)$ from (\ref{disk density distribution}),
$V_R(R,\theta)$ from (\ref{vr defn}), and $V_{\theta}(R,\theta)$ from
(\ref{vtheta defn}). To first order in the small quantities
\fpert, \epsdisk, and their derivatives:
\begin{equation}
  \frac{R}{R_d} \epsdisk(R) = \fpert(R) \left[ a_{12}(R) \left(1 -
    \frac{R}{R_d} + \frac{R}{V_c} \frac{\diffd V_c}{\diffd R} \right)
    - a_{32}(R) \right].
  \label{epsdisk fpert a12 reln}
\end{equation}
The neglected second-order terms induce
small $m=4$ perturbations in the disk;
see Appendix~\ref{second order terms section} for details.%
\footnote{We have also omitted the term
in equation (\ref{epsdisk fpert a12 reln}) that is proportional
to $R\> \diffd(\fpert\, a_{12})/\diffd R$; however, its
effect is negligible.}
Equation~(\ref{epsdisk fpert a12 reln})
provides us with a relationship between the radial profile of
the potential (embodied in $a_{12}$, $a_{32}$ and $V_c$), the strength of
the perturbation in the net potential (\fpert), and
the ellipticity of the disk (\epsdisk).

\subsection{Disk perturbation potential}\label{disk potential section}
The ellipticity of the disk generates an $m=2$ perturbation to the
disk potential.
We calculate this as follows:
\begin{eqnarray}
 \Phi^{\mathrm{disk}}_{\mathrm{pert}} \equiv
  \Phi^{\mathrm{disk}} - \Phi_0^{\mathrm{disk}} =
    -G \sum_{m=-\infty}^{\infty} \exp(i m \theta)
  \int_0^\infty J_m(k R)\, \exp(-k\, |z|)\, \diffd k
\nonumber\\
  \int_0^\infty J_m(k R')\, R'\, \diffd R'
  \int_0^{2\pi} \left[\Sigma(R',\theta') - \Sigma_0\, \exp(-R'/R_d)\right]
    \, \exp(-i m \theta')\, \diffd \theta'
\label{full disk integral}
\end{eqnarray}
(\citealp{BT}, eq. 2P-8).
We restrict ourselves to the plane $z=0$, corresponding to an
infinitely thin disk.

For small perturbations, the perturbed surface density is
\begin{equation}
  \Sigma(R',\theta') - \Sigma_0\, e^{-R'/R_d} \approx 
    \Sigma_0\, e^{-R'/R_d}
    \frac{R'}{R_d} \frac{\epsdisk(R')}{2} \cos 2\theta'.
\label{sigma pert}
\end{equation}
Substituting (\ref{sigma pert}) into (\ref{full disk integral}),
we note that the integral over $\diffd \theta'$ vanishes except
when $m=\pm 2$.
Since $\int_0^{2\pi} \cos 2\theta'\, e^{\pm i 2 \theta'}\, \diffd\theta'
=\pi$, $e^{ix}+e^{-ix}=2\cos x$, and $J_2(x) = J_{-2}(x)$, we find
\begin{equation}
 \Phi^{\mathrm{disk}}_{\mathrm{pert}}(R,\theta)
    = -\pi G\, \Sigma_0 \cos 2 \theta
    \int_0^{\infty} J_2(k R)\,\diffd k
  \int_0^{\infty} J_2(k R')
  R'\, \exp\left(-\frac{R'}{R_d}\right)
    \frac{R'}{R_d} \epsdisk(R')\, \diffd R'.
\end{equation}
We can express $(R'/R_d)\, \epsdisk(R')$ in the final integral in terms
of \fpert, $a_{12}$, $a_{32}$, and $V_c$ using (\ref{epsdisk fpert a12 reln}):
\begin{eqnarray}
 \Phi^{\mathrm{disk}}_{\mathrm{pert}}(R,\theta)
    = -\pi G\, \Sigma_0 \cos 2 \theta
    \int_0^{\infty} J_2(k R)\,\diffd k 
  \int_0^{\infty} J_2(k R')\, R'\, \exp\left(-\frac{R'}{R_d}\right)
\nonumber\\
  \fpert(R')
  \left[a_{12}(R') \left( 1 - \frac{R'}{R_d} +
   \frac{R'}{V_c(R')}\frac{\diffd V_c(R')}{\diffd R'}\right) -  
   a_{32}(R')\right]
   \, \diffd R'.
\end{eqnarray}

Because we do not know \textit{a priori} the net perturbation
$\fpert(R)$, we cannot immediately evaluate these integrals.
However, if we can find a function
$\fpert^{\mathrm{proxy}}$ whose form is similar to $\fpert$,
i.e. if $\fpert/\fpert^{\mathrm{proxy}}$
is a slowly-varying function of $R$, then we can approximate
the potential as:
\begin{eqnarray}
 \Phi^{\mathrm{disk}}_{\mathrm{pert}}(R,\theta) \approx
   -\frac{\fpert(R)}{\fpert^{\mathrm{proxy}}(R)} \pi G\, \Sigma_0 \cos 2 \theta
    \int_0^{\infty} J_2(k R)\,\diffd k
  \int_0^{\infty} J_2(k R')\, R'\, \exp\left(-\frac{R'}{R_d}\right)
\nonumber\\
    \fpert^{\mathrm{proxy}}(R')
  \left[a_{12}(R') \left( 1 - \frac{R'}{R_d} +
   \frac{R'}{V_c(R')}\frac{\diffd V_c(R')}{\diffd R'}\right) -  
   a_{32}(R')\right]
   \, \diffd R'.
\label{eq fproxy integral}
\end{eqnarray}
Note that $a_{12}(R')$ and $a_{32}(R')$ depend implicitly on
$\fpert^{\mathrm{proxy}}(R')$ through its derivative.
A first approximation can be obtained by setting
$\fpert^{\mathrm{proxy}} = \fpert^{\mathrm{halo}}$, whose values
have been tabulated from (\ref{eq fperthalo}).
Because the integral over $R'$ is independent of $R$
and the integral over $k$ is independent of $R'$,
the integrals can be evaluated independently on fine grids
of $k$ and $R$, respectively.
Using this technique, we calculate
\begin{eqnarray}
  \eta(R) \equiv \frac{\pi G\, \Sigma_0}{\fpert^{\mathrm{proxy}}(R)}
    \int_0^{\infty} J_2(k R)\, \diffd k
  \int_0^{\infty} J_2(k R')\, R'\, \exp\left(-\frac{R'}{R_d}\right)
\nonumber\\
    \fpert^{\mathrm{proxy}}(R')
  \left[a_{12}(R') \left( 1 - \frac{R'}{R_d} +
   \frac{R'}{V_c(R')}\frac{\diffd V_c(R')}{\diffd R'}\right) -  
   a_{32}(R')\right]
   \, \diffd R'
\label{eq etar defn}
\end{eqnarray}
at each grid point $R_i$. The perturbation potential due to the disk
is then given by
\begin{equation}
  \Phi^{\mathrm{disk}}_{\mathrm{pert}}(R,\theta) = -\fpert(R)\,
    \eta(R)\, \cos 2\theta.
\label{eq phi disk resp}
\end{equation}
Expressed in this form, the meaning of $\eta(R)$ becomes clear: it
is the magnitude of the disk response to a unit perturbation
in the potential.

\subsection{Self-consistent solution}\label{self-con soln section}
For clarity, we repeat here the important equations:
\begin{eqnarray}
\label{eq phipert vs phi0}
  \Phi_{\mathrm{pert}}(R,\theta) &=& \Phi_0(R)\, \fpert(R)\, \cos 2\theta \\
   &=&
    \Phi_{\mathrm{pert}}^{\mathrm{halo}}(R,\theta) +
    \Phi_{\mathrm{pert}}^{\mathrm{disk}}(R,\theta),
\end{eqnarray}
\begin{equation}
  \Phi_{\mathrm{pert}}^{\mathrm{halo}}(R,\theta) = \fpert^{\mathrm{halo}}(R)\,
    \Phi_0(R)\, \cos 2\theta,
\end{equation}
(see eqs. \ref{eq fpert defn} and \ref{eq fpert breakdown}), and
\begin{equation}
  \Phi_{\mathrm{pert}}^{\mathrm{disk}}(R,\theta) = -\fpert(R)\,
    \eta(R)\, \cos 2\theta,
\label{eq fpert repeat}
\end{equation}
(repeated from eq. \ref{eq phi disk resp}).
The physical interpretation of these equations is that
the disk response is proportional to the net perturbation \fpert,
which is itself the sum of the disk response and the imposed halo
perturbation. The disk response is opposite in sign to the halo
perturbation,
so \fpert\ in the self-consistent solution must
be reduced with respect to the imposed halo perturbation.

The self-consistent solution can be obtained by collection
equations (\ref{eq phipert vs phi0}) through (\ref{eq fpert repeat}):
\begin{equation}
  \fpert(R) = \fpert^{\mathrm{halo}}(R) \frac{1}{1 + \eta(R)/\Phi_0(R)}.
\end{equation}
In other words, the response of the disk causes the overall potential
perturbation to be reduced by a factor
of $1 + \eta(R)/\Phi_0(R)$. All terms on the right hand side have
been tabulated at the grid points $R_i$, resulting in a trivial
evaluation of $\fpert(R)$.

\begin{figure}
\plotone{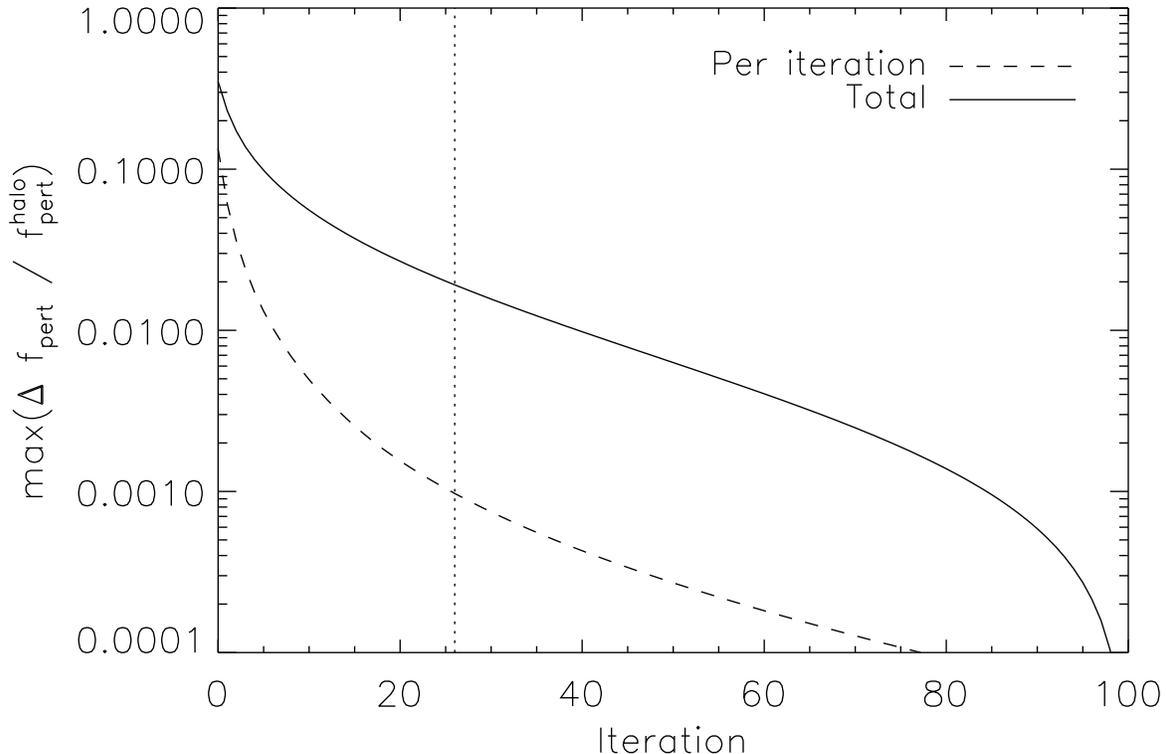}
\caption{\label{convergence figure}%
Maximum change in the solution for $\fpert/\fpert^{\mathrm{halo}}$
per iteration (dashed line) and maximum
difference between the solution at a given iteration and the true
solution, assumed to have converged after $100$ iterations (solid line),
for the fiducial halo and $3\times 10^9~\Msun$ disk
of \S~\ref{disk mass results section}.
The vertical dotted line indicates the point
where our convergence criterion is achieved.}
\end{figure}

Armed with this new estimate of \fpert, we can reexamine equation
(\ref{eq etar defn}), substitute $\fpert^{\mathrm{proxy}}=\fpert$,
and calculate a new value of $\eta(R)$ and therefore of $\fpert(R)$.
We repeat this procedure until the maximum change
between iterations in the quantity $\fpert/\fpert^{\mathrm{halo}}$,
which is a robust indicator of the relative error in \fpert,
is less than $10^{-3}$ at all radii; this is typically achieved
in $20$--$30$ iterations. We have confirmed for some specific cases
that our solution agrees
to within $\sim 2\%$ of the true solution (assumed to have converged after
a very large number of iterations) at all radii, and to much higher precision
at most radii
(see Figure~\ref{convergence figure}).
Adopting a stricter convergence criterion has no effect on our results.
The relatively large number of iterations is required in order to
accurately capture the sharp feature where $\fpert \rightarrow 0$
that is seen in the solutions (see \S~\ref{results section}).

Given \fpert, the disk ellipticity \epsdisk\ can be calculated as a function
of radius directly from equation (\ref{epsdisk fpert a12 reln}) and the
forms of the closed orbits can be calculated from equations
(\ref{closed orbit R}) -- (\ref{closed orbit Vtheta}). This provides a
complete description of the disk.

\section{Results}\label{results section}
In this section, we demonstrate how the disk dilutes
the elliptical potential of the halo,
and give examples of the net ellipticity
induced in the disk.
In \S~\ref{disk mass results section}, we demonstrate the main
features of the disk-halo systems using a halo with constant
axis ratios,
while in \S~\ref{varying halo parameters section} we investigate
how these results are affected by varying disk and halo
parameters such as the halo concentration, axis ratio, disk
scale length, and run of halo axis ratio with radius.

\subsection{Response for various disk masses}
\label{disk mass results section}
We demonstrate the main features of our models
using a fiducial triaxial
NFW halo with mass $M_{200}=10^{12}~\Msun$, axis ratios
$b/a=0.8$ and $c/a=0.6$, and a concentration $c_{200}=12$.%
\footnote{$M_{200}$ and $c_{200}$ refer to the mass and
concentration relative to the radius $r_{200}$, defined such that
the mean density within $r_{200}$ is $200$ times the critical
density. We assume the Hubble parameter $h=0.7$.}
These values are typical for galaxy-sized dark matter halos
in cosmological simulations \citep{allgood-etal06}.
The disk rotation axis is aligned with the minor axis of the halo,
in agreement with the orientation of the halo angular momentum
in simulations \citep{bs05-alignment}.
All disks have radial scalelengths $R_d=3.0~\mathrm{kpc}$, with
masses that range from zero up to $10^{11}~\Msun$.

\begin{figure}
\plotone{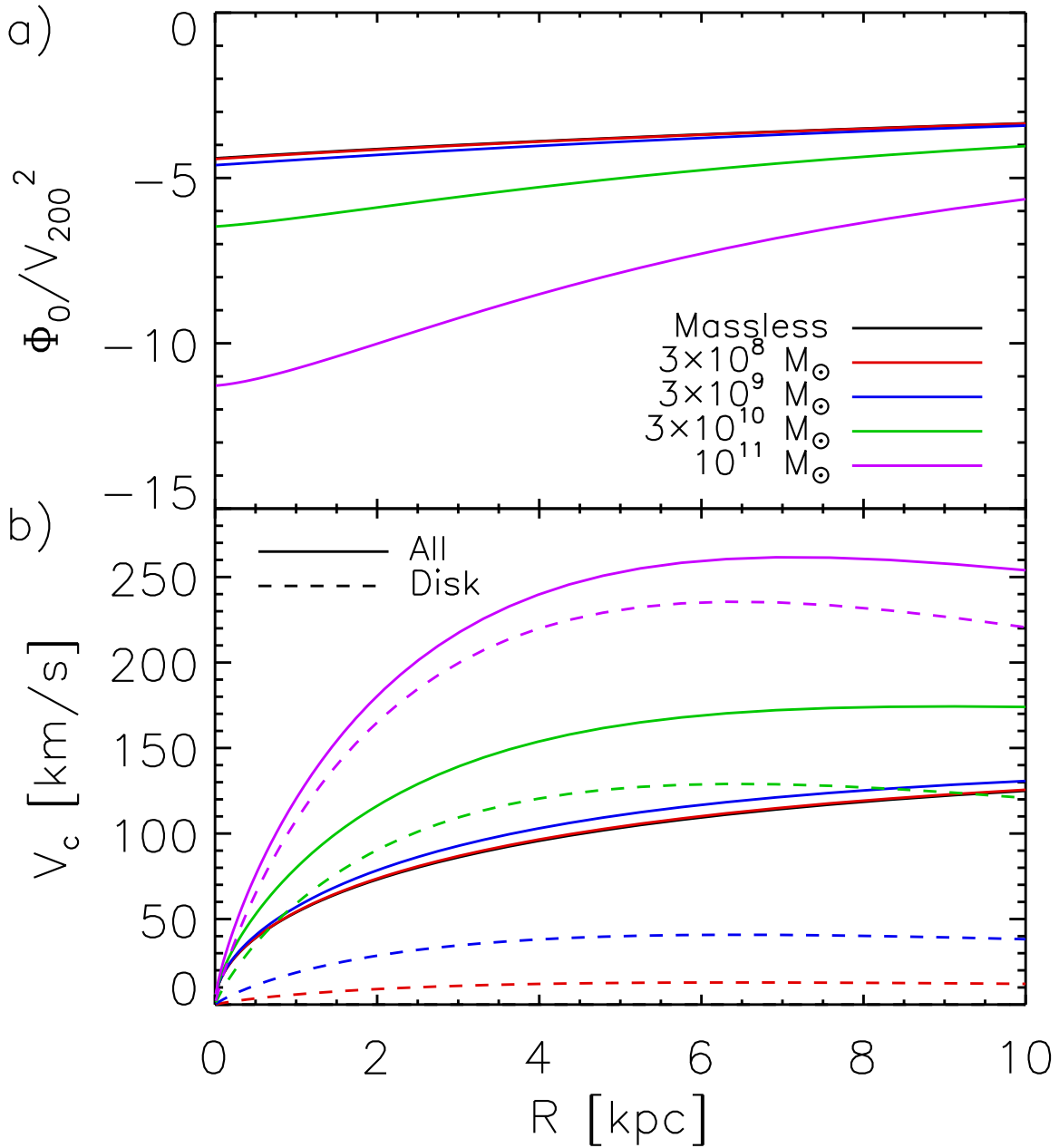}
\caption{\label{phi0 figure}%
\textit{(a)} Axisymmetric component of the potential, $\Phi_0(R)$, for
massless disks (black), and disks of mass $3 \times 10^8~\Msun$ (red),
$3 \times 10^9~\Msun$ (blue), $3 \times 10^{10}~\Msun$ (green),
and $10^{11}~\Msun$ (purple) within the fiducial halo
of \S~\ref{disk mass results section}.
\textit{(b)} Circular velocity curve within the unperturbed potential.
The disk contribution to the rotation curve is denoted with dashed
lines. Colors are the same as in \textit{(a)}.}
\end{figure}

\begin{figure}
\plotone{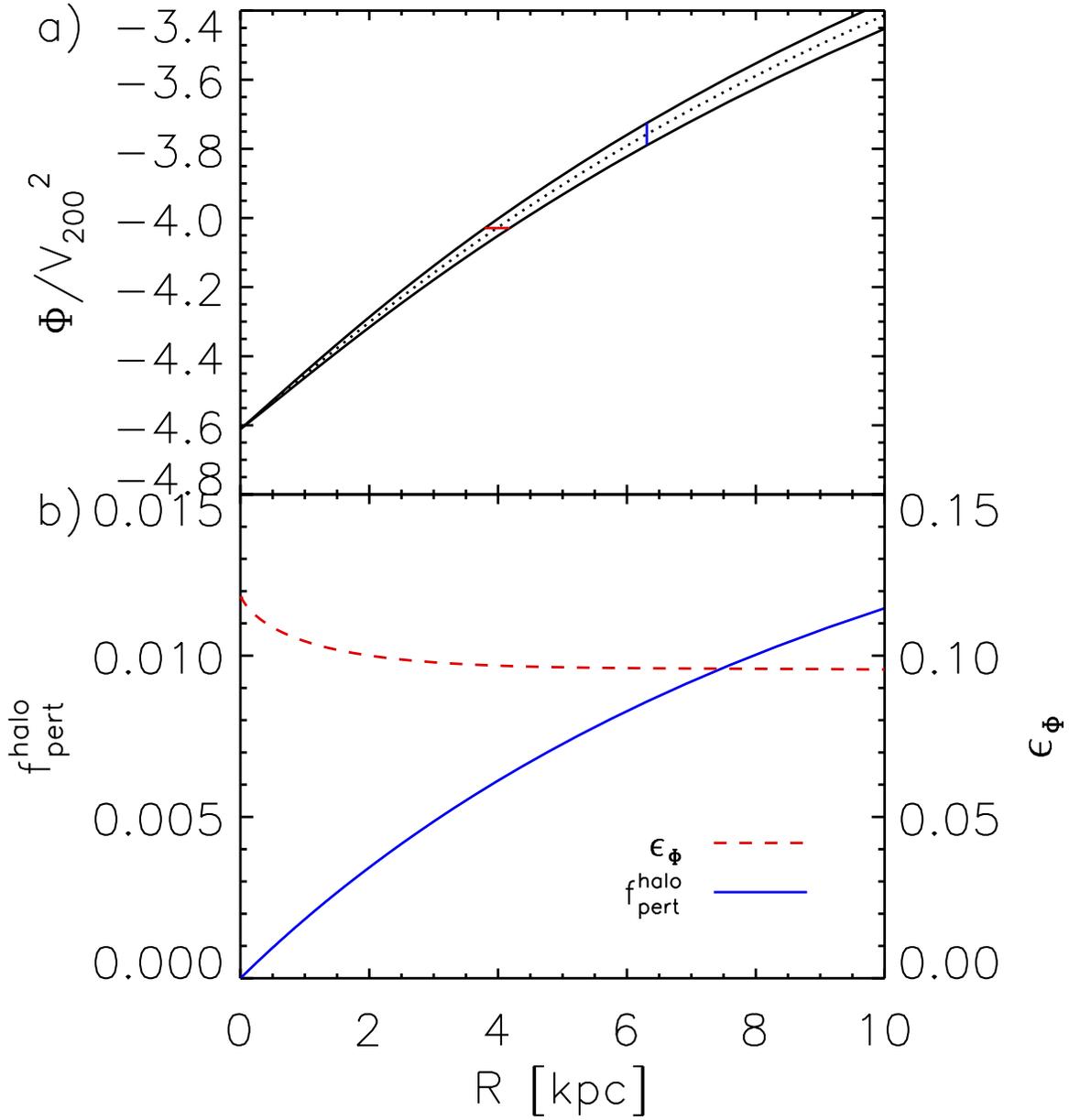}
\caption{\label{phixy figure}%
\textit{(a)}
The solid curves indicate the
potential along the $x$ (lower) and $y$ (upper) axis
for the disk of mass $3\times 10^9~\Msun$ within the fiducial
halo of \S~\ref{disk mass results section}. The dotted line indicates
the axisymmetric component of the potential. The horizontal and vertical
lines demonstrate how the ellipticity of the isopotential surfaces
and the magnitude of the perturbation respectively are calculated.
\textit{(b)}
The curves indicate the magnitude of the perturbation, \fpert\ (solid), and 
ellipticity of the potential, $\epsilon_{\Phi}$ (dashed) for the same
halo as in (a).}
\end{figure}

\begin{figure}
\epsscale{0.6}
\plotone{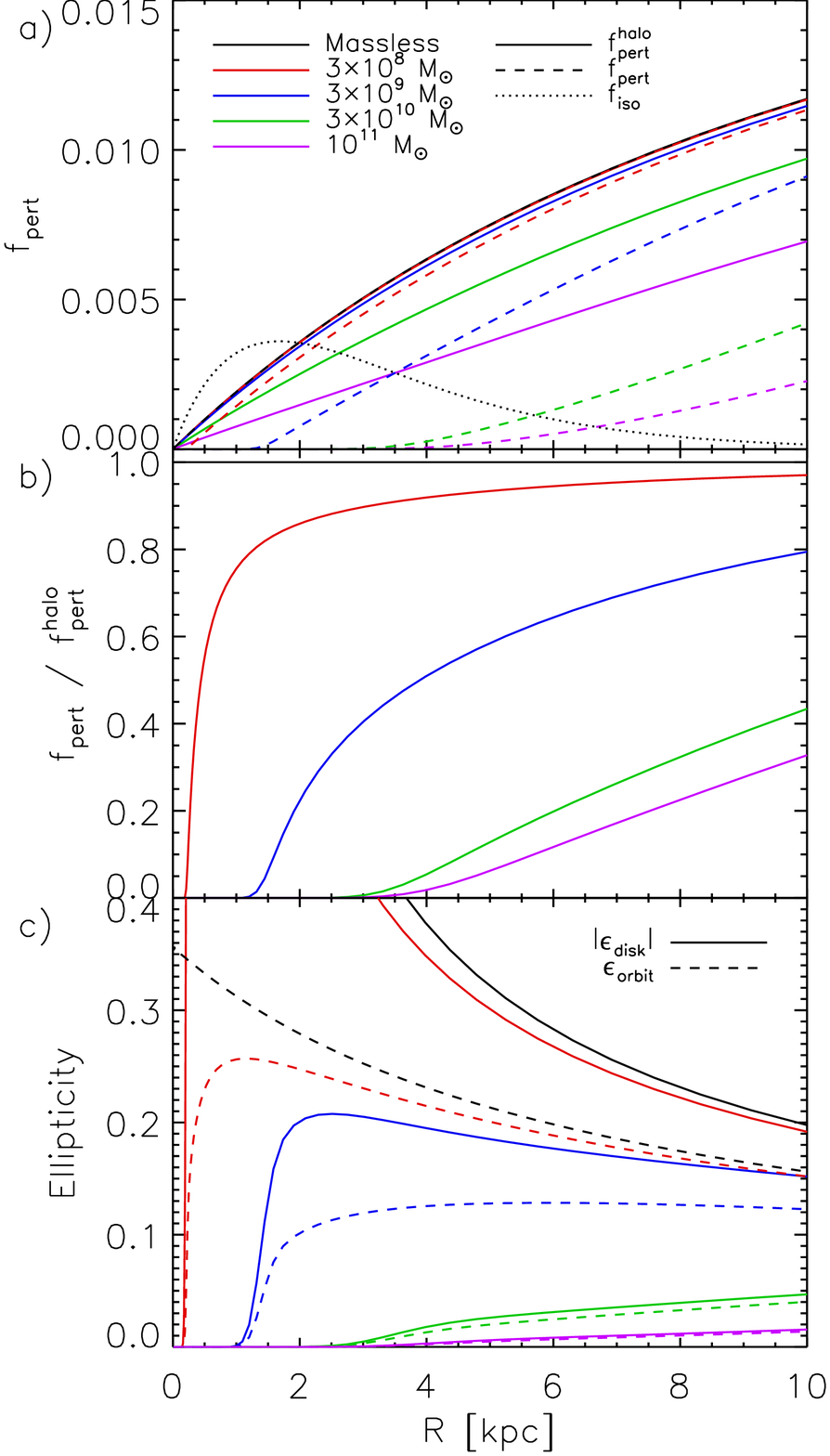}
\caption{\label{disk mass figures}%
\textit{(a)}
Magnitude of the elliptical perturbation in the potential due to
the triaxiality of the halo ($\fpert^{\mathrm{halo}}$, solid lines),
and the net perturbation after
including the self-consistent response of the disk
(\fpert, dashed lines). Different disk masses are indicated by different
colors as in Figure 1.
The dotted line shows the form of \fpert\ proposed by \citetalias{hn06}
to produce a rotation curve mimicking a cored isothermal
density profile.
\textit{(b)}
Ratio by which the input perturbation $\fpert^{\mathrm{halo}}$ becomes
diluted due to the self-consistent response of the disk.
\textit{(c)}
Ellipticity of the disk isodensity contours ($|\epsdisk|$, solid lines;
note that $\epsdisk$ is negative for positive \fpert\ in the sign
convention we have chosen) and of orbits within the disk
($\epsilon_{\mathrm{orbit}}$, dashed lines).}
\end{figure}

Figure~\ref{phi0 figure} demonstrates how the axisymmetric component
of the potential, $\Phi_0(R)$, and the rotation
curve, $V_c(R)$, vary as a function of disk mass.
The potential is given in units of $V_{200}^2 \equiv G\, M_{200}
/ r_{200}$.
For disk masses less then $3 \times 10^9~\Msun$, the halo dominates
the axisymmetric component of the potential at all radii, with the disk
becoming increasingly more important with increasing disk mass beyond
this.
The non-axisymmetric component of the potential is demonstrated in
Figure~\ref{phixy figure} for the $3 \times 10^9~\Msun$ disk. Although
the ellipticity of isopotential surfaces, $\epsilon_{\Phi}$, rises
to small radii, the magnitude of the potential perturbation,
$\fpert^{\mathrm{halo}}$, must vanish at small radii because
$\Phi_x(R=0) = \Phi_y(R=0)$ while $\Phi_0(R=0)$ reaches a finite
value.

Figure~\ref{disk mass figures}a demonstrates both
the initial halo perturbation ($\fpert^{\mathrm{halo}}$, solid lines)
and the net perturbation after including the
self-consistent response of disk (\fpert, dashed lines).
Note that even though the halo is identical in each case,
$\fpert^{\mathrm{halo}}$ is lower for higher mass disks because
the disk contributes to $\Phi_0$.

\citetalias{hn06} proposed that a suitable perturbation
\begin{equation}
 \fpert = f_{\mathrm{iso}}(R) \approx a\, x\, e^{-x/b}
\end{equation}
(with $x \equiv R/r_s$, $a=0.1$, and $b=0.098$) would cause 
the rotation curve of a perturbed NFW profile to mimic that of a
cored isothermal profile. We denote this as a dotted line in
Figure~\ref{disk mass figures}a.
The form of this perturbation is very different from the
form of the perturbation that we find to be induced by a triaxial halo
of uniform axis ratio,
particularly once the self-consistent response of the disk is
taken into account.

Figure~\ref{disk mass figures}b demonstrates the degree to which the initial
halo perturbation $\fpert^{\mathrm{halo}}$ is diluted by the
self-consistent response of the disk. This is equal to
$\fpert/\fpert^{\mathrm{halo}}$ and is determined from
$1/(1 + \eta(R)/\Phi_0(R))$.
The ellipticity in the potential vanishes in the central region
where $\eta(R) \gg \Phi_0(R)$.
Even for a negligible disk mass, the potential in the
innermost region is circularized.  This region is larger
for more realistic disks, which have a
significant impact on the potential out to several disk scale lengths.
As the disk mass increases, the form of the disk dominates
both $\eta(R)$ and $\Phi_0(R)$, and therefore this
function approaches an asymptotic form.

Comparison between Figure~\ref{disk mass figures}b
and the equivalent figs.~2 and 3 of
\citetalias{jog00} reveals dramatically different behavior at small radii:
in \citetalias{jog00}, the ``reduction factor'' reaches a minimum at $1.42~R_d$
($=4.26$~kpc for $R_d=3.0$~kpc) and then rises to unity,
while in Figure~\ref{disk mass figures}b
it falls monotonically to vanish at small radii.
This is a direct result of the radial variation of \fpert\ 
in a physically realistic elliptical halo.
Because $\eta(R)$ depends inversely on
\fpert\ (see eq.~\ref{eq etar defn}), which must vanish at small
radii, $\eta(R)$ must dominate over $\Phi_0(R)$ in the inner
regions and the potential must become completely circularized.

In Figure~\ref{disk mass figures}c we plot the ellipticity of
the disk isodensity contours ($\epsdisk$, solid lines) and of the
orbits within the disk ($\epsilon_{\mathrm{orbit}} \equiv \fpert\, a_{12}$,
dashed lines).
For a massless disk, we recover the
results of \citetalias{hn06} that the ellipticity rises towards
the center of the halo.
However, the presence of a massive disk changes the situation
dramatically. Because $\fpert/\fpert^{\mathrm{halo}}$ vanishes at small
radii (Figure~\ref{disk mass figures}b), the equilibrium disk is
axisymmetric at small radii. Even very low mass disks, which contribute
negligibly to $\Phi_0$,
still respond strongly enough to the elliptical potential
to cause an important change in the behavior at small radii.
We also note that the ellipticity of disk isophotes are always
greater than ellipticity of orbits within the disk.

\subsection{Varying disk and halo parameters}
\label{varying halo parameters section}
\begin{figure}
\epsscale{0.6}
\plotone{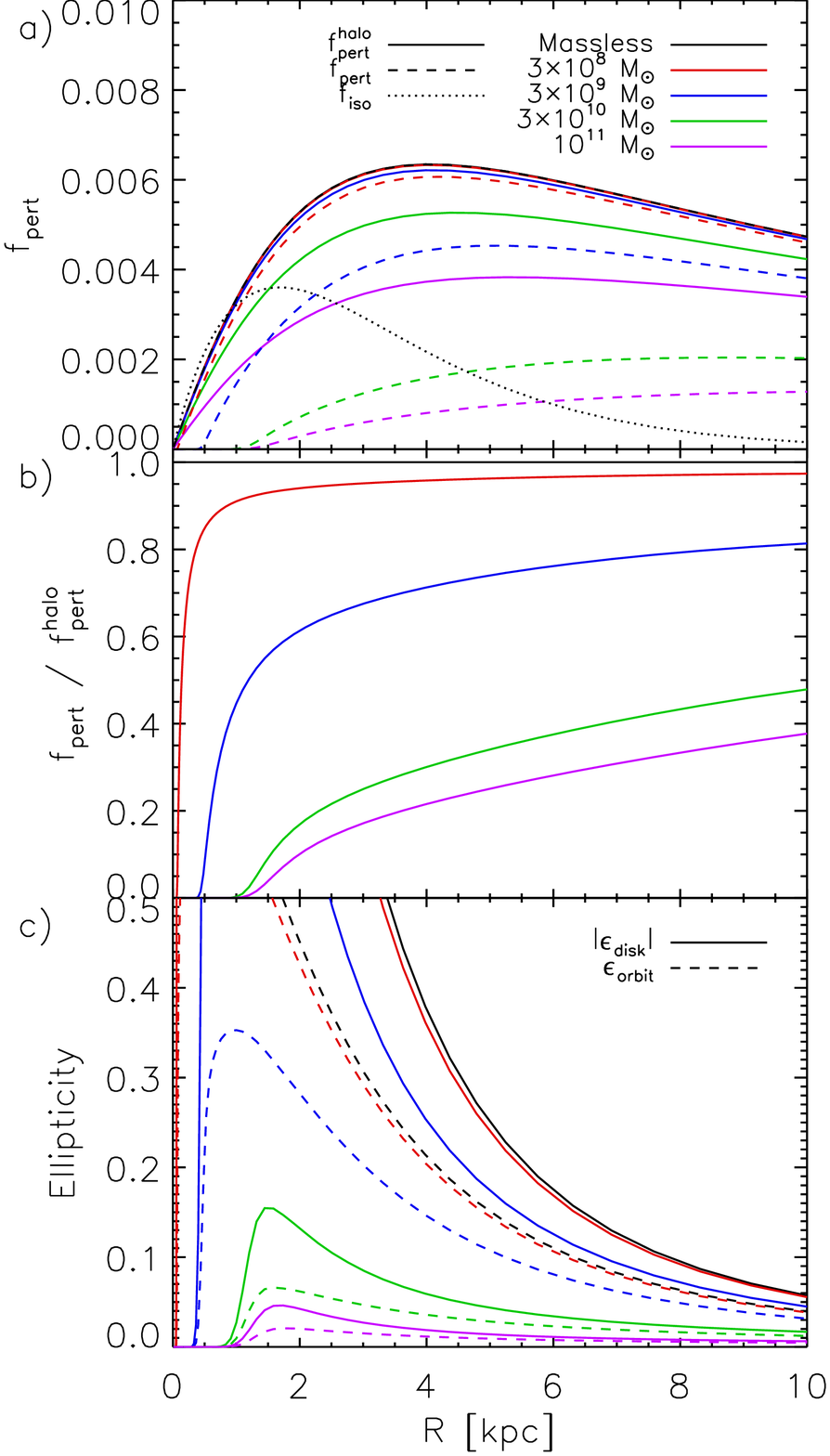}
\caption{\label{hns figures}%
As in Figure~\ref{disk mass figures}
for disks in a halo potential with the same
radial variation of its axis ratios as halo G4 of \citetalias{hns07}.}
\end{figure}
\citetalias{hns07} found that the radial variation of the shape of
the potential of cosmological $N$-body halos
is not consistent with self-similar isodensity
contours \citep[see also][]{js02,bs05-alignment}.
They found instead that the isopotential axis ratios are well fit
by the following function:
\begin{equation}
\log( \frac{b}{a} \mbox{ or } \frac{c}{a} ) =
  \alpha \left[ \tanh\left( \gamma \log\frac{r}{r_{\alpha}} \right)
  - 1 \right].%
\label{HNS equation}
\end{equation}
We have recomputed the self-consistent response of disks of varying
mass in a potential of this form, with the values of the parameters
taken from halo~G4 of \citetalias{hns07},
which has a very similar mass and concentration
to the halo used in \S~\ref{disk mass results section}.
The results are shown in Figure~\ref{hns figures}.
As noted by \citetalias{hns07},
the perturbation in this case contains a peak at intermediate radius
and is much more similar to the form of $f_{\mathrm{iso}}$ required
by \citetalias{hn06} to fit LSB rotation curves, although
unlike $f_{\mathrm{iso}}$ the
perturbation remains more prominent to large radius.  However, in many
cases LSB rotation curves are only measured out to radii of a few kpc,
so it is less clear what the required form of $f_{\mathrm{iso}}$ is at
larger radii.
The disks have less effect on the perturbation
than in \S~\ref{disk mass results section}; in particular,
the radius inside which they circularize the potential is reduced,
resulting in significantly more elliptical orbits at $1$--$3$~kpc
than for the equivalent disks in a halo with constant
axis ratios.  
Disk masses higher than $\sim 3\times 10^9~\Msun$
reduce the prominence of the peak in the perturbation and shift
it to larger radii.
The peak in $\fpert^{\mathrm{halo}}$ could be moved to smaller
radius, and therefore brought further into agreement with
$f_{\mathrm{iso}}$, by reducing the $r_{\alpha}$ parameter in
equation~(\ref{HNS equation}); however, this is unlikely to
be a common situation, as G4 already has by far
the lowest $r_{\alpha}$ value of any of the halos studied
by \citetalias{hns07}.

The effect of the baryonic disk on the shape of the halo
is not yet well understood. Simulations
suggest that halos containing baryonic disks
are less elliptical than halos composed purely
of dark matter, and that the circularization of the halo occurs most
strongly at the center \citep{kazantzidis-etal04}. If the
intrinsic shape of the pure dark matter halo is well described by
the \citetalias{hns07} form,
which is most elliptical at the center, baryonic processes
may result in a situation more similar to the constant axis
ratio case of \S~\ref{disk mass results section}.
We therefore expect that the regions in which these models
differ most strongly, $1$--$3$~kpc, are also the regions where the unknown
effect of disk formation introduces the most uncertainty into our models.

\begin{figure}
\epsscale{0.6}
\plotone{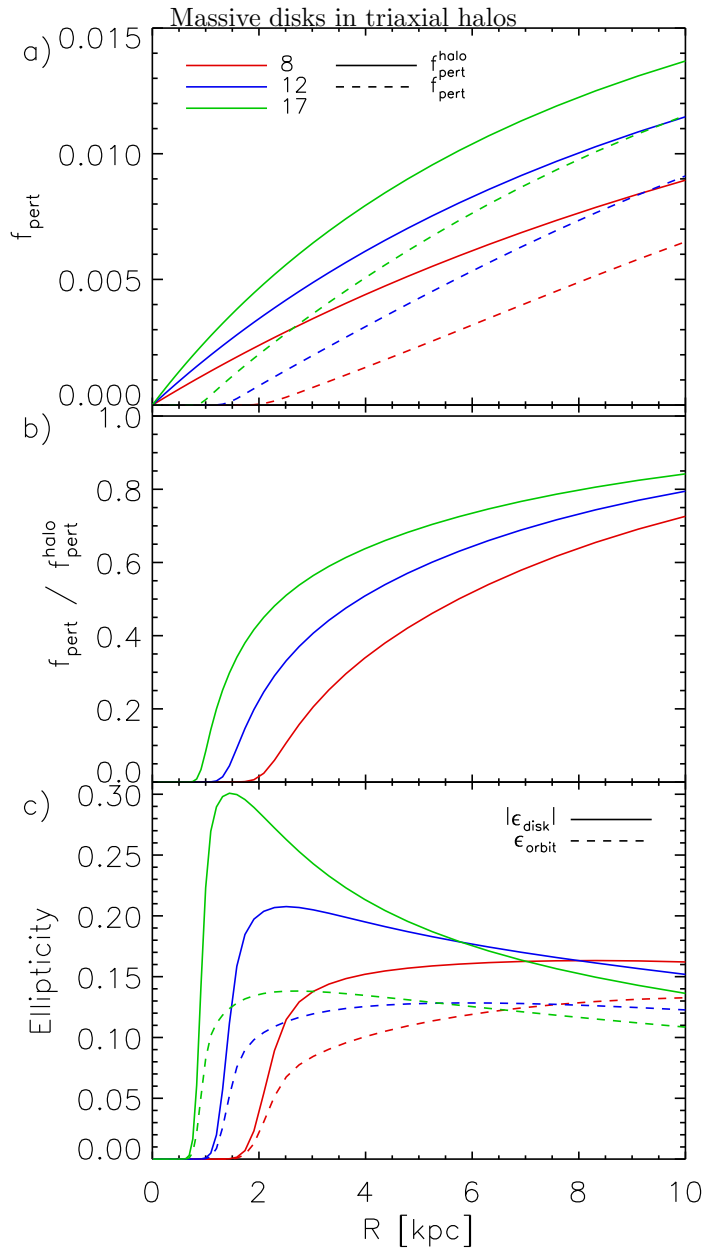}
\caption{\label{concentration figures}%
\textit{(a)} Magnitude of the net elliptical perturbation in the potential
(dashed lines) and the perturbation due to only the halo (solid lines)
for disks of mass $3 \times 10^9~\Msun$ in halos with
concentrations, $c_{200}=8$ (red), $12$ (blue), and $17$ (green).
The blue lines in these plots are identical to the blue lines
in Figure~\ref{disk mass figures}.
\textit{(b)} Ratio by which the input perturbation becomes
diluted due to the response of the disk.
\textit{(c)} Ellipticity of the disk isodensity contours
(solid lines) and of orbits within the disk (dashed lines).}
\end{figure}

\begin{figure}
\epsscale{0.6}
\plotone{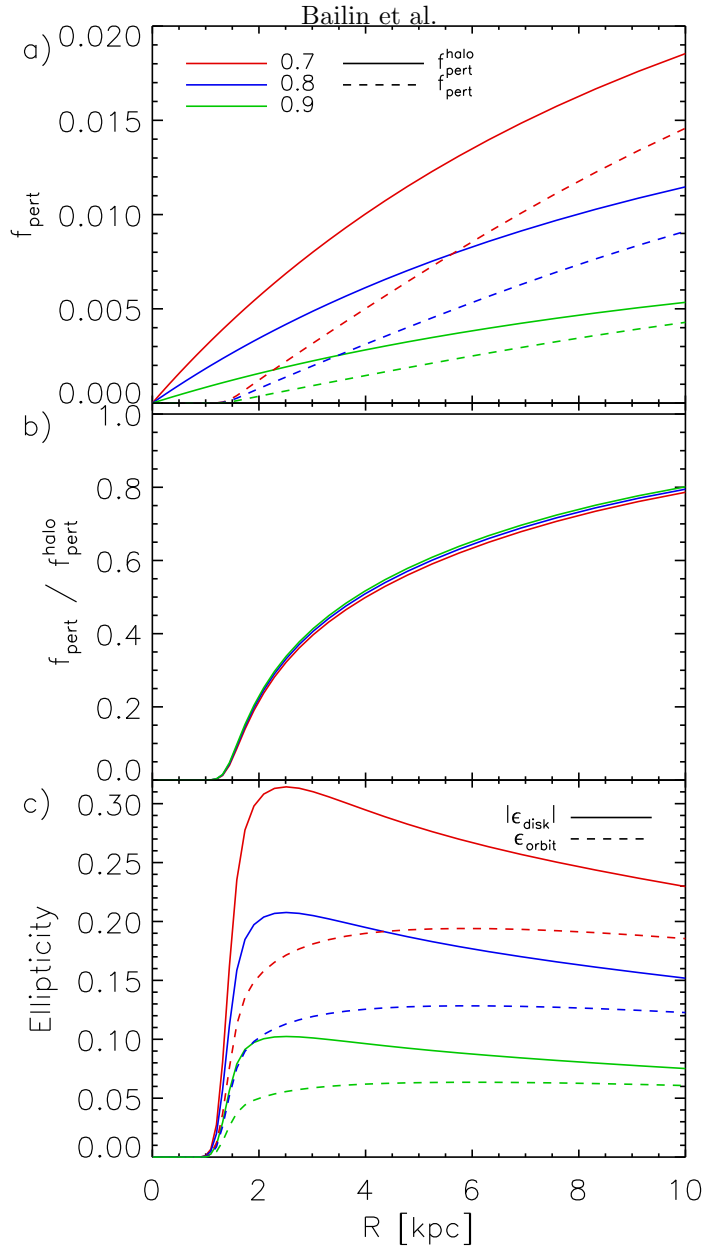}
\caption{\label{ba figures}%
As in Figure~\ref{concentration figures} for halos with
axis ratios $b/a=0.7$ (red), $0.8$ (blue), and $0.9$ (green).}
\end{figure}

\begin{figure}
\epsscale{0.6}
\plotone{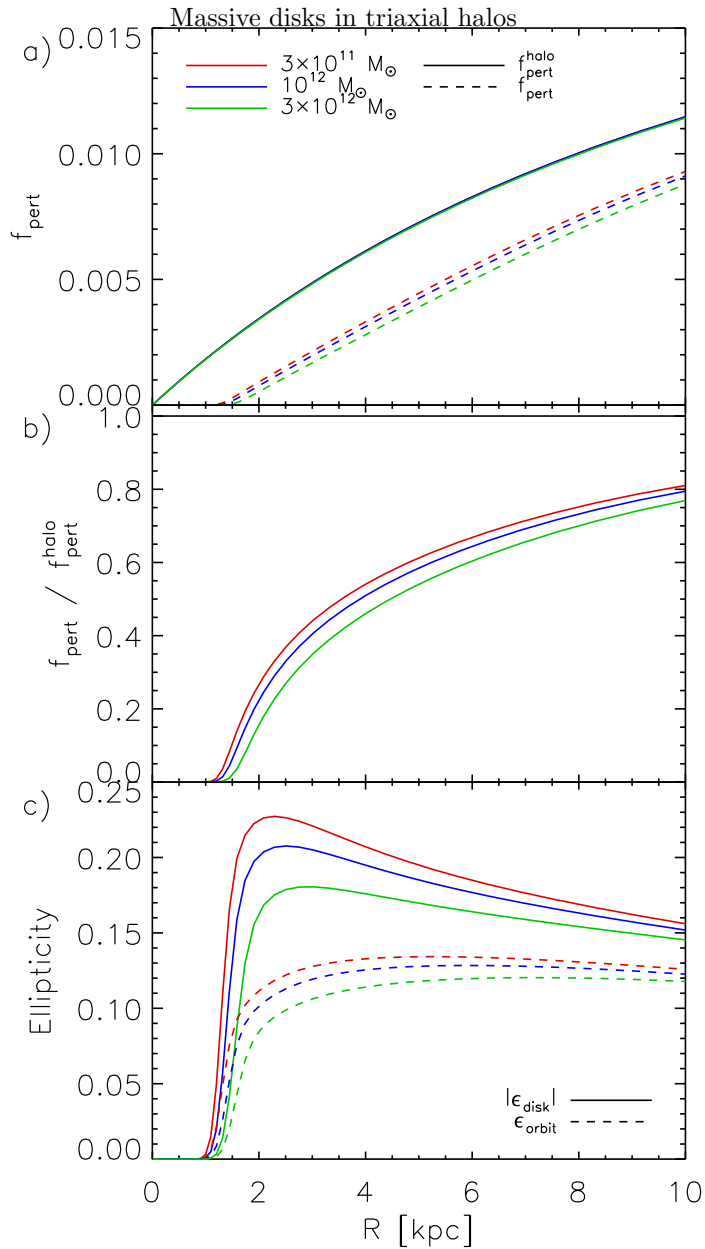}
\caption{\label{sys figures}%
As in Figure~\ref{concentration figures} for halos with
virial masses $M_{200}=3\times 10^{11}~\Msun$ (red),
$10^{12}~\Msun$ (blue), and $3\times 10^{12}~\Msun$ (green).
In order to facilitate
the comparison, the halo scale radius $r_s$ is kept constant
by varying $c_{200}$ in proportion to $r_{200}$, and the
ratio between the disk and halo mass is kept constant.}
\end{figure}

\begin{figure}
\epsscale{0.6}
\plotone{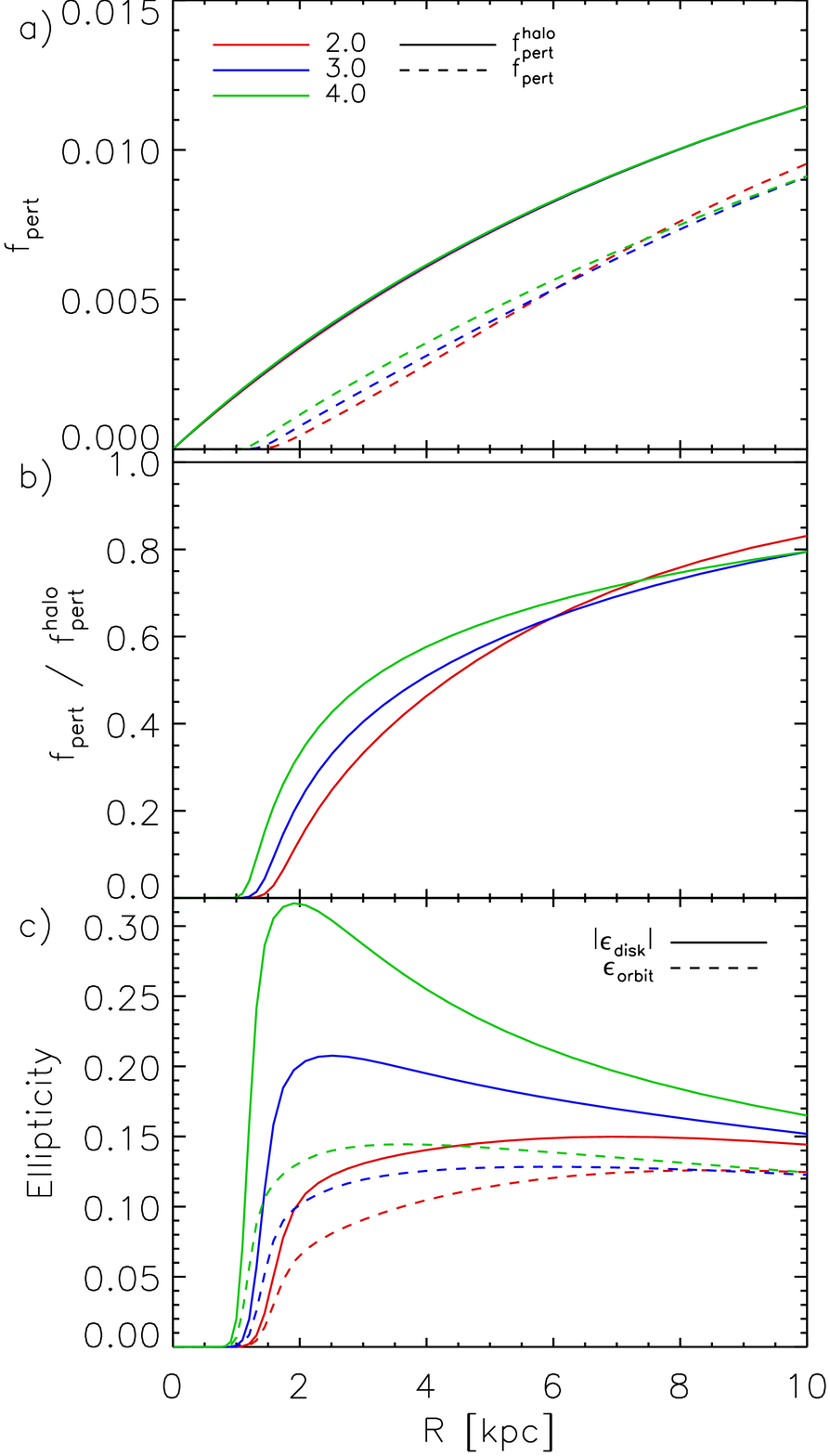}
\caption{\label{rd figures}%
As in Figure~\ref{concentration figures} for disks with
scale lengths $R_d=2.0$ (red), $3.0$ (blue), and $4.0$ (green).}
\end{figure}

In order to investigate how other properties of the halo
and disk affect our results,
we have recalculated the results of \S~\ref{disk mass results section}
(where the halo axis ratio was assumed to be constant with radius)
for the $3 \times 10^9~\Msun$ disk while
varying the halo concentration,
the $b/a$ axis ratio,
the halo mass, and
the disk scale length.
In more concentrated halos (Figure~\ref{concentration figures}),
the strength of the perturbation due to
the halo, $\fpert^{\mathrm{halo}}$, is larger.
The disk is also less able to dilute the perturbation in more
concentrated halos.
The axis ratio of the halo (Figure~\ref{ba figures}) has a strong effect
on the magnitude of the perturbation but has virtually
no effect on the degree to which the disk dilutes the perturbation.
In Figure~\ref{sys figures} we compare halos of different
virial mass, $M_{200}$.
In order to facilitate comparison between systems of different
mass, we have kept $r_s$ constant by varying $c_{200}$ in
proportion to $r_{200}$ and kept the relative mass of the
disk and halo constant.
We find that the mass of the halo has little effect on the
relative strength of the perturbation (either before or
after the disk is taken into account), but that the
resulting disk ellipticities are higher in lower mass systems.
Finally, although the equilibrium shape of the potential
is similar regardless of the disk scale length
(Figure~\ref{rd figures}),
a greater ellipticity is required to achieve this reduction in the
perturbation for less concentrated disks,
i.e.~those with larger scale lengths.

\section{Discussion}\label{discussion section}
\subsection{Impact of Triaxial Halos on the Cusp/Core Problem}
\begin{figure}
\plotone{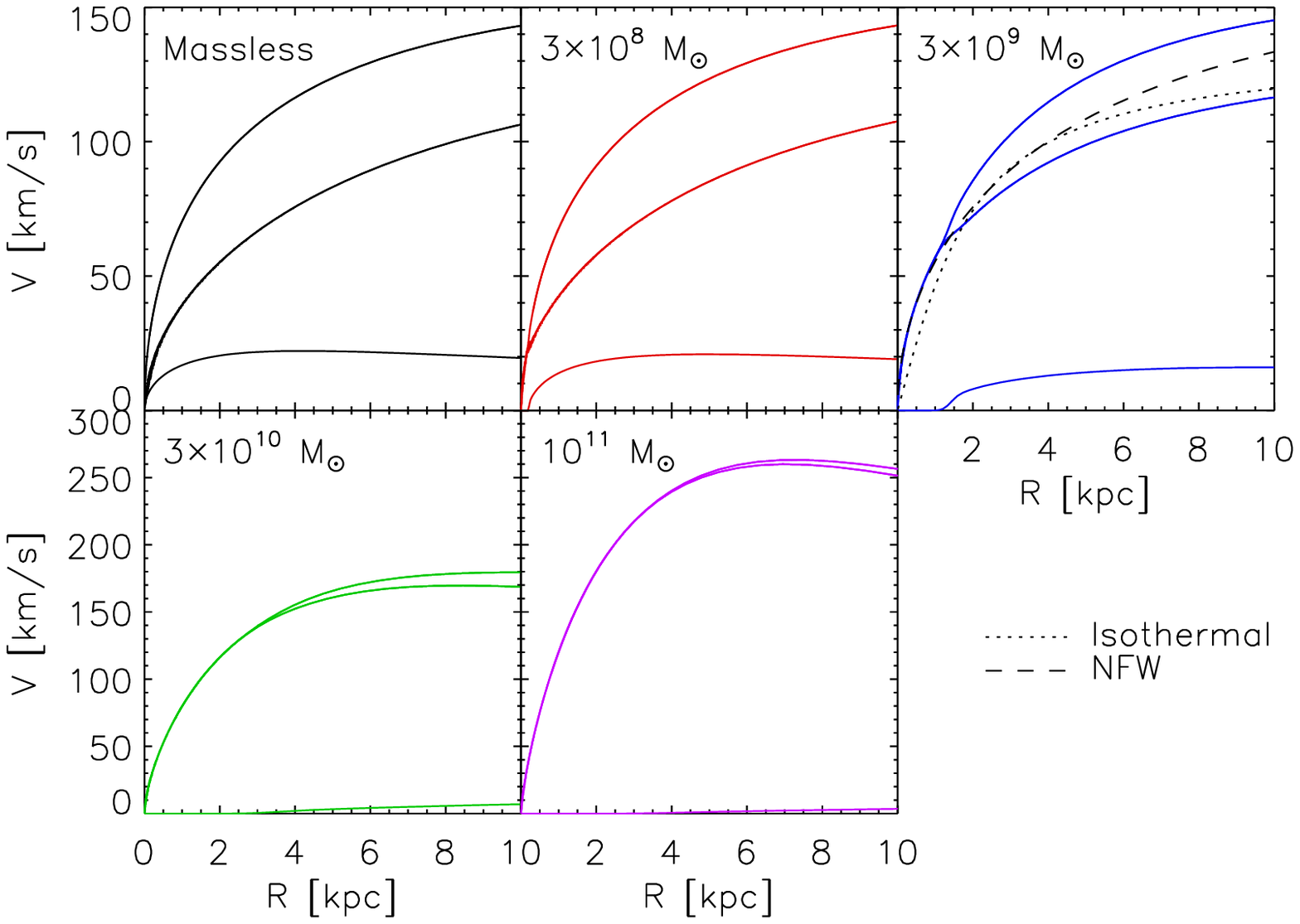}
\caption{\label{vmajmin figure}%
Azimuthal velocity along the halo major axis (top line) and minor
axis (middle line),
and the maximum radial velocity at each radius (bottom line)
for disks of varying mass in the halo of
\S~\ref{disk mass results section}.
NFW and isothermal rotation curves are shown
for reference in the top right panel.}
\end{figure}

\begin{figure}
\plotone{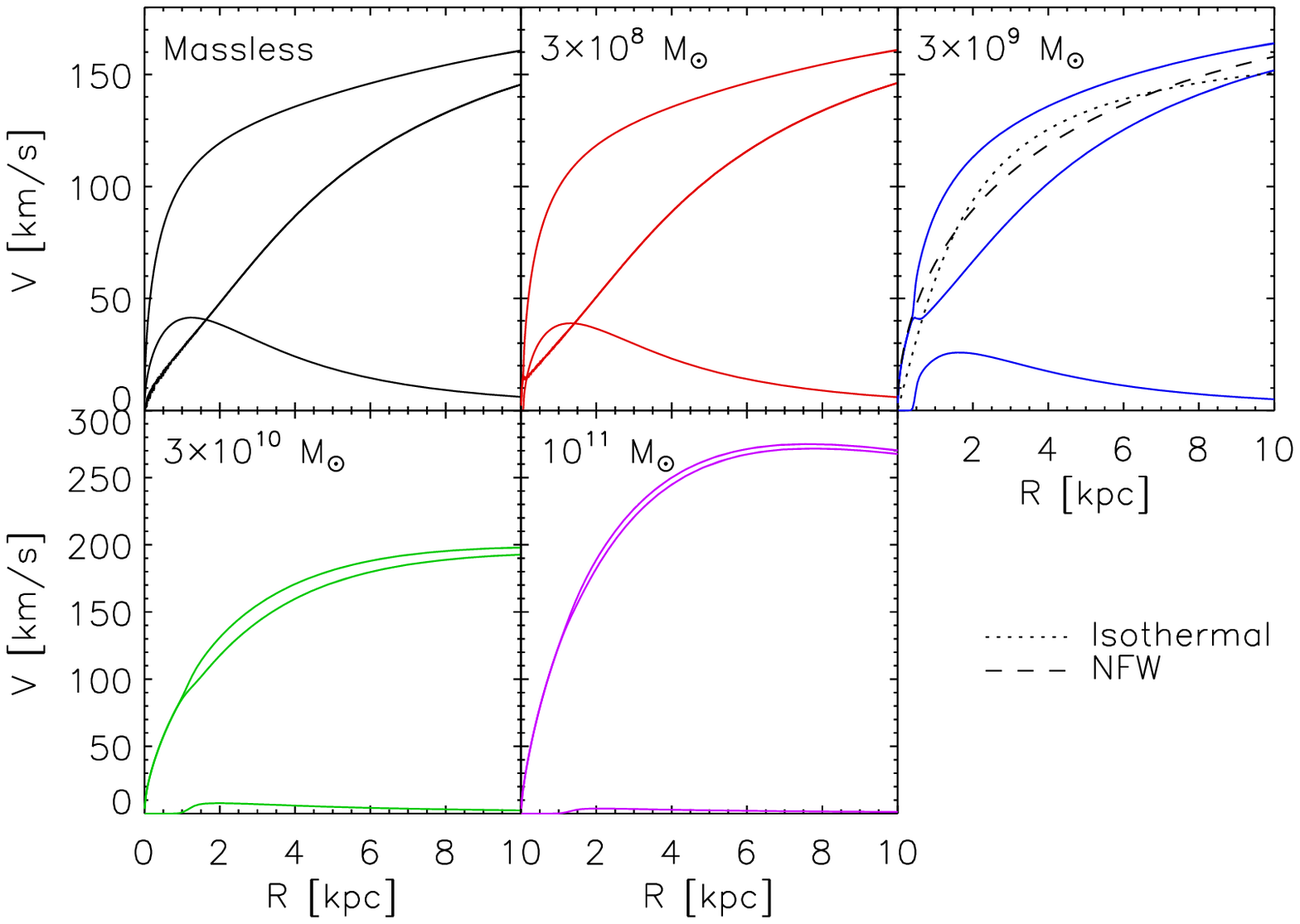}
\caption{\label{vmajmin hns figure}%
As in Figure~\ref{vmajmin figure}, but for the shape
of the halo potential found by \citetalias{hns07} for
their halo~G4.}
\end{figure}

In Figure~\ref{vmajmin figure}, we plot the azimuthal velocities along
the major and minor axes of the halo within the disk plane
for disks of different mass in the fiducial sample halo studied
in \S~\ref{results section}.
Full analysis of the velocity fields
of these disks will be presented in a future paper.
However, we note that none of these rotation curves shows the
linear rise characteristic of a constant density core,
as expected given the dramatic difference between the form
of \fpert\ in these disks and the form of $f_{\mathrm{iso}}$
required by \citetalias{hn06}.

If we use the radial variation of the shape of the halo potential proposed
by \citetalias{hns07} (Figure~\ref{vmajmin hns figure}),
for which $\fpert^{\mathrm{halo}}$ is much more similar to
$f_{\mathrm{iso}}$, we find that
for very low disk masses the azimuthal velocity along the
minor axis of the halo is characterized by a
much more gradual rise.  Observationally, such a rotation curve might
be interpreted as indicating a constant-density core in the dark
matter halo.
However, for disk masses above $3 \times 10^9~\Msun$ ($0.3\%$ of the
virial mass of the halo and just $1.8\%$ of the baryonic
mass of the system) the response of the disk removes this
feature from the center of the rotation curve.
Based on these results, we conclude that simple analyses of the 
shapes of halos are therefore not
sufficient to determine whether halo triaxiality can
reconcile LSB rotation curves with cuspy halo density profiles,
as suggested by \citetalias{hn06}; full analyses that take
into account the disk response are required.  Preliminary tests on
simulated velocity fields constructed using the results of this paper
do suggest that triaxiality can produce apparent constant-density
cores, in agreement with \citetalias{hn06}, but a more detailed
analysis including many halos and lines of sight is needed before 
comparing to the observational distribution of density profile 
slopes \citep{simon-etal05}.

We also plot the maximum radial velocity (amplitude of non-circular
motions) at each radius in
Figures~\ref{vmajmin figure} and \ref{vmajmin hns figure}.
Although the radial velocities are much smaller than
the azimuthal velocities in all but the lowest mass disks,
they are at a level that can be detected in observations
of two dimensional velocity fields. The magnitude of the
radial motions, which reach $5$--$35~\mathrm{km~s^{-1}}$ depending
on the disk mass and halo properties, are consistent with
the magnitude of radial motions found by
\citet{simon-etal05}, and reproduce the observed trend
for the radial motions to be negligible at small
radii and to only become important at larger radii.

\subsection{Comparison to Observed Disk Ellipticities}

It is interesting to compare the ellipticity of our model disks to
observed values.  Using two-dimensional velocity fields,
\citet{simon-etal05} found lower limits on the orbital ellipticities
ranging from $0$ up to $0.175$, similar to the orbital ellipticities
in our model disks.  We note that the orbital ellipticity throughout
most of the disk is determined by the mass of the disk
(Figure~\ref{disk mass figures}c), the ellipticity of the halo
(Figure~\ref{ba figures}c), and the variation of halo ellipticity with
radius (compare Figures~\ref{disk mass figures}c and \ref{hns
  figures}c), with very little dependence on the concentration of
either the halo (Figure~\ref{concentration figures}c) or the disk
(Figure~\ref{rd figures}c), or on the global mass of the system
(Figure~\ref{sys figures}c).  We therefore predict that galaxies with
large observed ellipticities such as \object{NGC 4605} either lie in halos that
are more triaxial than average near their center, or contain an
unusually low fraction of their mass in their disk.
The trends in intrinsic disk ellipticity that we predict may
be tested with further analysis of larger kinematic samples,
such as those presented by \citet{ganda-etal06}.

\citet{ryden06} found that the distribution of isophotal shapes
of galaxies in the 2MASS Large Galaxy Atlas \citep{2mass-lga},
as measured in the near-infrared $K_s$ band (which is a good tracer
of the stellar disk mass), is well fit if
the intrinsic disk ellipticity distribution is a truncated Gaussian
centered at $0.01$ with a width of $\sigma=0.26$. This corresponds
to a median ellipticity of $0.18$, with 68\%\ of disks having
ellipticities $0.05 \le \epsdisk \le 0.37$. For the disk
parameters we have studied, the
$18.8~\mathrm{mag~arcsec^{-2}}$ isophote at which her shapes
were measured corresponds to radii
of between $1.5$ and $8.5~\mathrm{kpc}$. Our models naturally
produce disk ellipticities in this range at these radii,
although the highest
ellipticities can only be produced by our least massive disks in
our most elliptical potentials.

Finally, we note that our models only take into account
ellipticity in the disk induced by the dark matter halo.
Central regions of the disk, which have high surface density
and sit in an axisymmetric potential, may be unstable to
bar formation \citep{bs06}. This can induce additional ellipticity
to the kinematic and photometric properties of disk galaxies
\citep[e.g.,][]{valenzuela-etal07}.

\section{Conclusions}\label{conclusions section}
We have presented a
computationally efficient method to self-consistently determine the dynamics
of massive disks in triaxial dark matter halos. Our work extends
the study of \citetalias{jog00} by allowing the perturbation
to the potential to vary with radius in an appropriate manner and by
allowing the ellipticity of the disk to vary with radius
self-consistently. These improvements result in qualitatively
different behavior for the ellipticity of disks at small radii:
\citetalias{jog00} found that disks counteract the halo
ellipticity most strongly at $1.42~R_d$
and have a negligible effect at small radii; in contrast, we find
that the effect of the disk increases monotonically
to small radii, completely circularizing the potential
in the innermost regions.

This self-consistent radially-varying response of the disk to the halo
perturbation must be taken into account when comparing the observed
kinematic and photometric properties of galactic disks to those
expected in triaxial dark matter halos, particularly for comparisons
at small radii.  When this response is calculated for plausible halo
values, model disks have ellipticities consistent with those
determined from observations of velocity fields and from isophotal
axis ratio distributions.  We also find that the radial variation of
the halo axis ratios has a significant impact on the disk structure.
Halos with axis ratios that vary with radius as suggested by
cosmological simulations produce much more elliptical orbits
in the inner disk than do halos with constant axis ratios,
resulting in potential perturbations similar to
the perturbation required to create apparent cores in galaxy density
profiles.  
Further analysis exploring in detail the conditions under
which core-like rotation curves might be obtained will be necessary to
determine if halo triaxiality can resolve the cusp/core problem.

\acknowledgements
JB thanks the Australian Research Council for financial support.  JDS
acknowledges the support of a Millikan Fellowship provided by Caltech.
BKG and CP gratefully acknowledge the support of the Australian
Research Council supported ``Commonwealth Cosmology Initiative'',
DP 0665574.
We thank Milo{\v s} Milosavljevi{\'c} and Darren Croton
for helpful conversations, and the referee, Chanda Jog,
for a very useful report.

\bibliography{../../masterref.bib}

\appendix
\section{Second-order terms and m=4 distortions}\label{second order terms section}
We have assumed that the halo perturbation and the
induced distortion in the disk are
completely described by the $m=2$ mode.
This a direct consequence of only including terms linear in
the small quantities \fpert, \epsdisk, and their derivatives.

The validity of this assumption can be tested by evaluating
the magnitude of the second-order terms that contribute to the
$m=4$ distortion in the disk. If we expand the potential as
\begin{equation}
  \Phi(R,\theta) = \Phi_0(R) \left(1 + f_2(R) \cos 2\theta +
    f_4(R) \cos 4\theta\right),
\end{equation}
the disk surface density as
\begin {equation}
  \Sigma(R,\theta) = \Sigma_0 \exp \left[ - \frac{R}{R_d}
    \left( 1 - \frac{\epsilon_2(R)}{2} \cos 2\theta
    - \frac{\epsilon_4(R)}{2} \cos 4\theta \right) \right],
\end{equation}
and include all second-order terms, then the $m=4$ distortion in the
disk, $\epsilon_4$, depends on terms of order
$f_4$, $f_2\, \epsilon_2$, and
$R\, f_2\, \diffd\epsilon_2/\diffd R$.%
\footnote{As in the case of the $m=2$ mode, there is also
a negligible term of order $R\, \diffd(f_4\, a_{14})/\diffd R$.}
\begin{figure}
\plotone{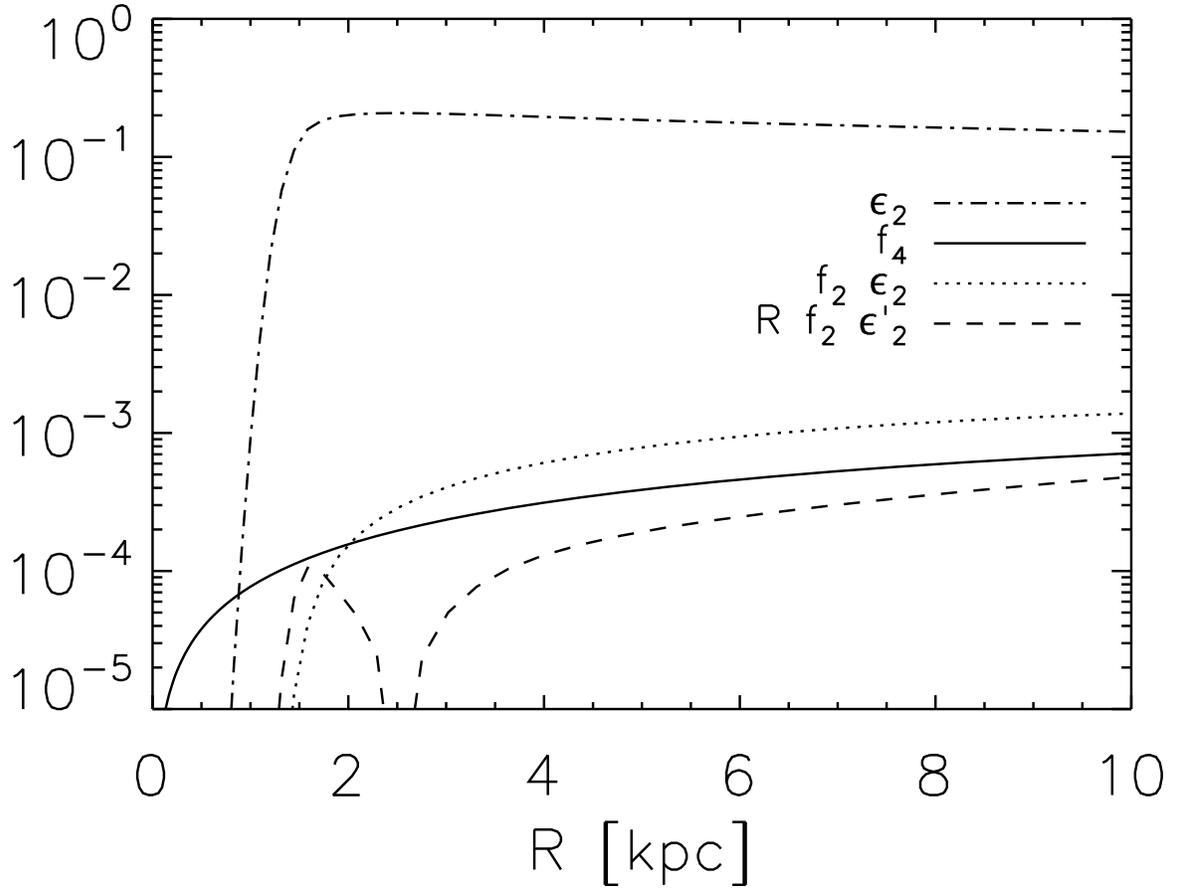}
\caption{\label{eps4 figure}%
Magnitude of terms contributing to the $m=4$ distortion
in the disk compared to $\epsilon_2$,
the magnitude of the calculated $m=2$ distortion
for a disk of mass $3 \times 10^9~\Msun$ in the fiducial halo
of \S~\ref{disk mass results section}.}
\end{figure}
Figure~\ref{eps4 figure} compares
the magnitude of these terms to the $m=2$ ellipticity for the
$3 \times 10^9~\Msun$ disk in the fiducial halo of
\S~\ref{disk mass results section}.
The higher-order terms are more than two orders of magnitude
smaller than the first-order terms over most of the disk,
and are also negligible in the central region where the first order terms
vanish,
validating our use of linear perturbation theory.

\end{document}